\documentclass[letter,twocolumn]{jpsj2} 

\def\runauthor{Hirono {\sc Kaneyasu}, Keita {\sc Kishigi} and Yasumasa {\sc Hasegawa}}

\title{Effects of Electron Correlation near Spin-Density-Wave on Angle Dependence of Magnetoconductivity}

\author{Hirono \textsc{Kaneyasu}\thanks{E-mail 
hirono@sci.u-hyogo.ac.jp}, Keita \textsc{Kishigi}$^{1}$
and Yasumasa \textsc{Hasegawa}}   
\inst{Department of Material Science, University of Hyogo, 
Kamigori, Akou, Hyogo 678-1297, Japan \\
$^{1}$Department of Education, Kumamoto University, Kumamoto 860-8555, Japan}

\abst{
The effect of electron correlation on the angle dependence of magnetoconductivity is studied in quasi-one-dimensional organic conductors. 
We investigated the effect on the basis of the momentum dependence of both quasi-particle's lifetime and velocity near the spin-density-wave (SDW).
We found that the momentum dependence of quasi-particle's lifetime mainly governs the angle dependence of one-dimensional axis magnetoconductivity. 
On the other hand, the change in velocity originating from the electron correlation gives the dominant effects on the angle dependence of the interchain axis magnetoconductivity. 
The effect of electron correlation is clarified from the association of the momentum dependences of lifetime and velocity with the momentum dependences of commensurate orbitals on a Fermi surface in magnetic fields.
}

\kword{magnetconductivity, electron correlation, angle dependence, one-dimensional organic conductor, spin-density-wave}

\begin{document}
\maketitle
Quasi one-dimensional (Q1D) organic conductors often 
show spin-density-wave (SDW) and field-induced SDW, which are caused by  strong electron correlation (EC).\cite{FISDW1,FISDW2}
In these systems, the angle-dependent magnetoresistance oscillation (AMRO) has been discovered as a profitable study for the fermiology of low-dimensional conductors (TMTSF)$_2$X.\cite{AMRO1,AMRO2,AMRO3}
Experimentally, minimums at magic angles are observed in both 1D axis and interchain magnetoresistances in the magnetic fields in a $y-z$  plane.\cite{AMRO1,AMRO2,AMRO3}
Here, the magic angles $\theta$ of a magnetic field are given from the relation of $\theta=\tan^{-1}(\frac{p}{q})$, where $p$ and $q$ are integers. \cite{AMRO1,AMRO2,AMRO3} 
The electron has commensurate routes on a Fermi surface (FS) at magic angles, whereas it has incommensurate routes at other angles.  
The strong angle dependence of interchain magnetoresistance was obtained in terms of the topology of the linearized dispersion in the $k_x$-direction on FS.\cite{OSADA1,OSADA2} 

On the other hand, the angle dependence of the 1D axis magnetoresistance is not obtained from the linearized FS because a 1D axis velocity has no momentum dependence. 
Although the angle dependence of 1D axis magnetoresistance is obtained from the unlinearized dispersion on the FS, it is much weaker than that observed experimentally.\cite{MAKI}
It is found that the angle dependence of 1D axis magnetoconductivity ($\sigma_{xx}$) appears when a spot of slow or fast velocity exists in  the momentum space of the 1D axis velocity.\cite{HASEGAWA} The result in this simple model indicates the possibility that the change in velocity near SDW causes the strong angle dependence in $\sigma_{xx}$.

As for another mechanism, the effect of Umklapp scattering\cite{UMKLAPP} near SDW was considered on the basis of the momentum dependence of scattering rate\cite{LEBED,CHAIKIN,YAKOVENKO,MCKENZIE} in the linearized dispersion on FS.
Lebed and Bak\cite{LEBED} have first pointed out that magnetoresistance in Q1D organic systems depends on the angles of the magnetic field by studying the effect of EC with impurities. 
They obtained, however, the peaks of magnetoresistivity at magic angles, instead of the dips (i.e., peaks of magnetoconductivity) observed in experiments. 
Zhelenznyak and Yakovenko\cite{YAKOVENKO} have studied the field angle  dependence of both 1D axis and interchain magnetoresistance in Q1D organic systems 
in a strong magnetic field on the basis of the semiclassical Boltzmann theory by taking account of the momentum-dependent scattering rate due to the topology of the Q1D FS. 
Yanase and Yamada\cite{YANASE} have studied the $c-$axis resistivity in quasi two-dimensional cuprate systems on the basis of the semiclassical Boltzmann theory, where both energy and the lifetime of quasi-particle are obtained using the second order perturbation theory in the on-site repulsion. 
Their study is performed only in zero magnetic field.

Here, 
we study $\sigma_{xx}$ and interchain magnetoconductivity ($\sigma_{zz}$) in the Q1D organic systems in the strong magnetic field tilted in the $y-z$ plane on the basis of the semiclassical Boltzmann theory by considering the effect of EC on both lifetime and energy (and its derivative, the velocity of quasi-particle) 
in a random phase approximation (RPA). 
The effect of EC is treated in the full momentum dependence with the unlinearized dispersion. 
Because spin susceptibility is expected to be strongly affected by EC in the Q1D organic systems, 
which are close to SDW instability, 
the effect of EC is analyzed using the more appropriate RPA than the second order perturbation theory.

\begin{figure}[!ht]
\begin{center}
\includegraphics[width=0.47\textwidth]{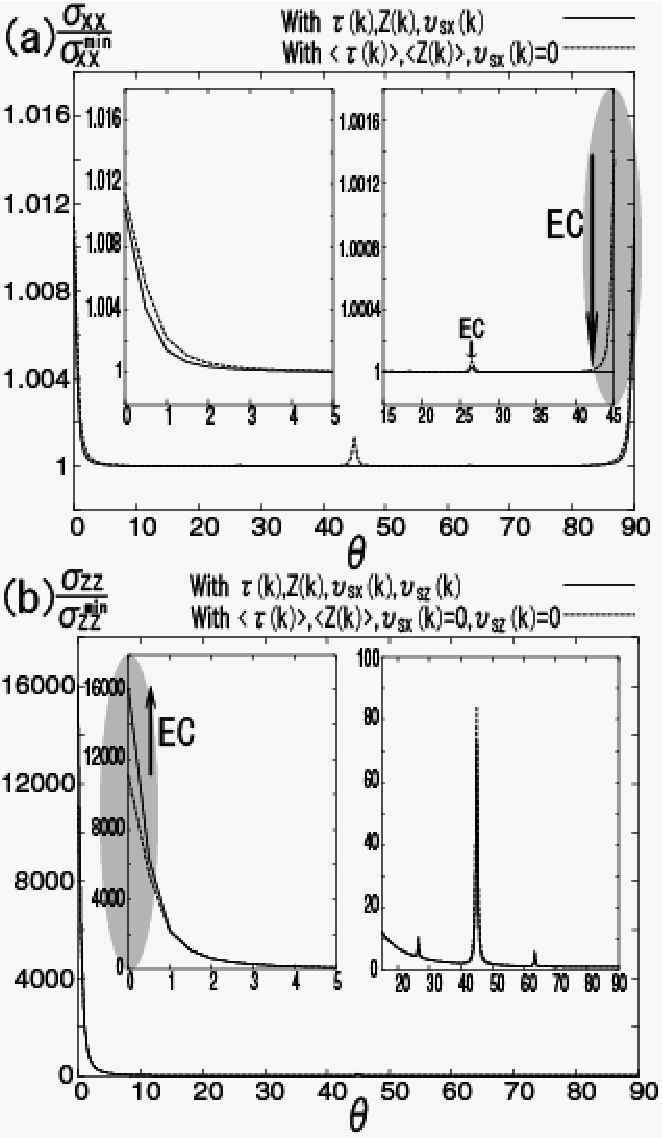}
\end{center}
\caption{(a) Angle dependence of $\sigma_{xx}$. (b) Angle dependence of $\sigma_{zz}$. 
The solid lines mean the magnetoconductivities with using the momentum dependent 
$\tau({\bf k})$, 
$Z({\bf k})$ and  
$\upsilon_{sx}({\bf k})$ 
(and $\upsilon_{sz}({\bf k}) in (b))$. 
The dash lines indicate the magnetoconductivies 
in the case that we neglect the momentum dependence of $\tau({\bf k})$, $Z({\bf k})$ and $\upsilon_{sx}({\bf k})$ 
(and $\upsilon_{sz}({\bf k})$ in (b)) by 
using
$\langle\tau({\bf k})\rangle$, 
$\langle Z({\bf k})\rangle$, 
and $\upsilon_{sx}({\bf k})=0$ 
(and $\upsilon_{sz}({\bf k})=0$ in (b)). 
}
\label{fig1}
\end{figure}
\begin{figure}[!ht]
\begin{center}
\includegraphics[width=0.47\textwidth]{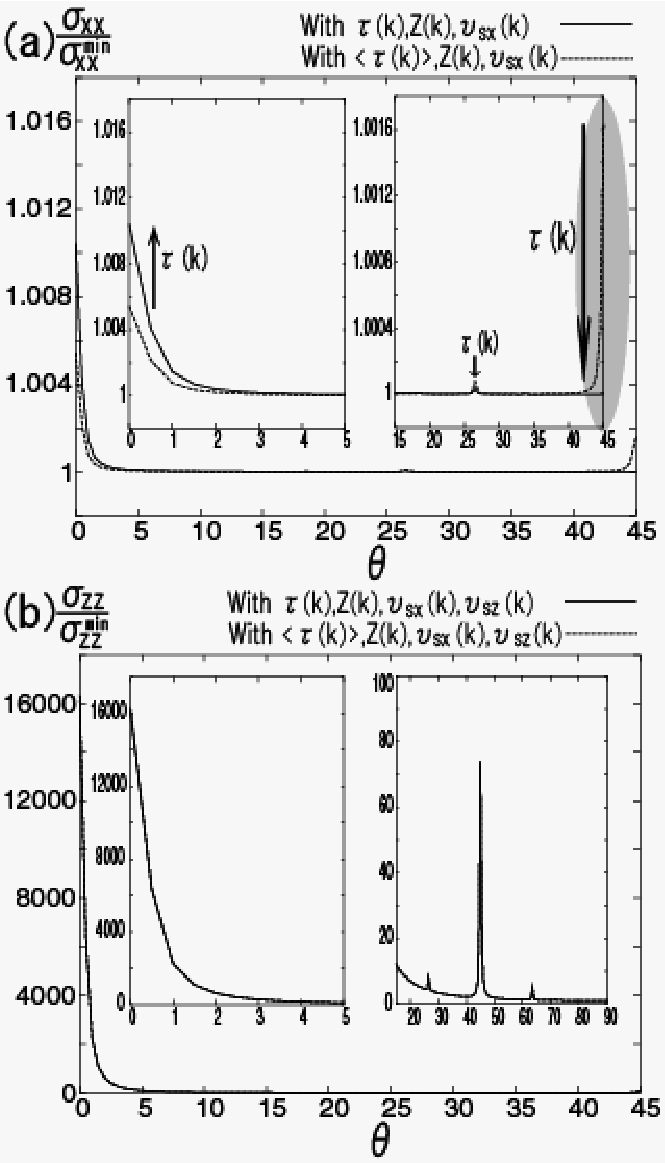}
\end{center}
\caption{(a) Angle dependence of $\sigma_{xx}$. (b) Angle dependence of $\sigma_{zz}$. 
The solid lines mean the magnetoconductivities with using the momentum dependent 
$\tau({\bf k})$, 
$Z({\bf k})$, 
$\upsilon_{sx}({\bf k})$ 
and $\upsilon_{sz}({\bf k})$. 
The dash lines indicate the magnetoconductivities 
in the case that the momentum dependence of $\tau({\bf k})$  is neglected by using $\langle\tau({\bf k})\rangle$. 
In $\sigma_{zz}$ in (b), two lines are overlapped and indistinguishable on the scale.
}
\label{fig2}
\end{figure}

As for formulation, the dispersion of a tight-binding model is given by 
$E_{\bf 0}({\bf k})=-2t_a\cos k_x-2t_b\cos k_y-2t_c\cos k_z,$
where the nearest-neighbor hopping parameters are $t_a=1.0$ and $t_b=t_c=0.1$ for the simplified Q1D FS of organic conductors. We use $t_a$ as the unit of energy.
The realistic parameters for (TMTSF)$_2$X are $t_b=0.1$ and $t_c=0.03$.
However, we use the symmetric parameter $t_b=t_c=0.1$
 for the saving of three-dimensional mesh in momentum space.
Electrons are set to be quarter-filled.
The self-energy $\Sigma({\bf k}, \omega_{\rm n})$ is obtained on the basis of RPA with respect to the on-site interaction $U$ in the Hubbard model. 
$\Sigma({\bf k}, \omega_{\rm n})$ is given by
\begin{equation*}
\Sigma({\bf k}, \omega_{\rm n})
=\frac{T}{N}\sum_{{\bf k'}, m}
U^2[
\frac{3}{2}
\chi_s({\bf k'}, \omega_{\rm m})
+\frac{1}{2}
\chi_c({\bf k'}, \omega_{\rm m})
\end{equation*}
\vspace{-4mm}
\begin{equation}
\hspace{3.1cm}-\chi_0({\bf k'}, \omega_{\rm m})
]
G_0({\bf k-k'}, \omega_{\rm n-m}).
\end{equation} 
Here, spin and charge susceptibilities are respectively 
$\chi_s({\bf k'},\omega_{\rm m})
=\frac{\chi_0({\bf k'}, \omega_{\rm m})}
{1-U\chi_0({\bf k'}, \omega_{\rm m})}$
and
$\chi_c({\bf k'}, \omega_{\rm m})
=\frac{\chi_0({\bf k'}, \omega_{\rm m})}
{1+U\chi_0({\bf k'}, \omega_{\rm m})}$
with 
$\chi_0({\bf k'}, \omega_{\rm m})
=-\frac{T}{N}
\sum_{{\bf k}, n}
G_0({\bf k+k'}, \omega_{\rm n+m})
G_0({\bf k}, \omega_{\rm n}),$
where $N$ is the number of sites and $\omega_n=\pi T(2 n+1)$ is the fermion Matsubara frequency with temperature $T$.
The quasi-particle's velocity
\begin{equation}
\upsilon_i({\bf k})=\upsilon_{0i}({\bf k})+\upsilon_{si}({\bf k})
=\frac{\partial E_0({\bf k})}{\partial k_i}+\frac{\partial Re\Sigma^R({\bf k}, E^*)}{\partial k_i}, 
\end{equation}
the quasi-particle's lifetime 
\begin{equation}
\tau({\bf k})=-\frac{1}{Im\Sigma^R({\bf k}, E^*)},
\end{equation}
and 
the renormalization factor 
\begin{equation}
Z({\bf k})=\frac{1}{1-\frac{\partial Re\Sigma^R({\bf k}, E)}{\partial E}|_{E=E^*}}
\end{equation}
are given on FS.
Here, $i$ indicates $x$, $y$ and $z$. $\Sigma^R({\bf k}, E^*)$ and $E^*({\bf k})$ indicate a retarded self-energy and a quasi-particle's energy on FS, respectively. The angle dependence of magnetoconductivity is calculated from the Boltzmann transport theory\cite{ASCROFT} and the equation of motion in magnetic field $B$, 
$\sigma_{ij}(\theta)=\frac{e^2}{4 \pi^3}
\int
-\frac{\partial f(E({\bf k}(0)))}{\partial E}|_{E=E^*}
\tilde{\upsilon}_{i}({\bf k}(0))$
$\int_{-\infty}^0
\tilde{\upsilon}_j({\bf k}(t))$
$\exp[-\int_t^0
\frac{dt'}{\tilde{\tau}({\bf k}(t'))}]
dtd{\bf k}(0).$
Here, 
$\tilde{\upsilon}_{i}({\bf k}(t))$ and $\tilde{\tau}({\bf k}(t))$ are $\tilde{\upsilon}_{i}({\bf k}(t))=Z({\bf k}(t))\upsilon_{i}({\bf k}(t))$ 
and 
$\tilde{\tau}({\bf k}(t))=\frac{\tau({\bf k}(t))}{Z({\bf k}(t))}$, 
respectively. 
$f(E({\bf k}))$ is the Fermi distribution function.
$-\frac{\partial f(E({\bf k}(0)))}{\partial E}|_{E=E^*}$ is the density of states on FS. 
We use the relation given by 
$-\frac{\partial f(E({\bf k}(0)))}{\partial
E}|_{E=E^*}=\frac{1}{\tilde{\upsilon}_x({\bf k}(0))}$ 
in the Q1D system.
$\sigma_{xx}$ ($i=j=x$) and $\sigma_{zz}$ ($i=j=z$) are
\begin{equation}
\sigma_{xx}(\theta)
=\frac{e^2}{4 \pi^3}
\int 
\int_{-\infty}^0 
\tilde{\upsilon}_x({\bf k}(t))
\exp[-\int_t^0 
\frac{dt'}{\tilde{\tau}({\bf k}(t'))}]
dtd{\bf k}(0)
\end{equation}
and 
\begin{equation*}
\sigma_{zz}(\theta)
=\frac{e^2}{4 \pi^3}
\int 
\frac{1}{\tilde{\upsilon}_{x}({\bf k}(0))}
\tilde{\upsilon}_z({\bf k}(0))
\int_{-\infty}^0 
\tilde{\upsilon}_z({\bf k}(t))
\end{equation*}
\vspace{-2mm}
\begin{equation}
\hspace{3.1cm}\times
\exp[-\int_t^0
\frac{dt'}{\tilde{\tau}({\bf k}(t'))}]
dtd{\bf k}(0).
\end{equation} 
The equation of motion is
$\hbar\frac{d{\bf k}(t)}{dt}=-e\tilde{\bf v}({\bf k}(t))\times{\bf B},$
where magnetic field ${\bf B}$ is in the $y-z$ plane and 
${\bf B}=(0, B\sin\theta, B\cos\theta)$.
In the numerical calculation, 
we divide the first Brillouin zone into $64^3$ momentum meshes
and take the cut-off $N_{f}$ = 128 for Matsubara frequency $\omega_n$.
The bandwidth $2W$ ($W\sim 2.2$) is a necessary
range of $\omega_n$ for reliable calculations, that is 
$W< \pi T N_{f}$. To satisfy the condition, 
the conductivity should be calculated in the region with $T>0.0055t_a$.
We choose the parameters as $T=0.008$, $U=1.98$ and $B\langle\tau({\bf k})\rangle=150$. 
Here, $\langle ...  \rangle$ means the average on FS. The SDW state is stabilized in the region of $U>3.1$ where $\chi_s(k)$ diverges.

\begin{figure}[!ht]
\begin{center}
\includegraphics[width=0.48\textwidth]{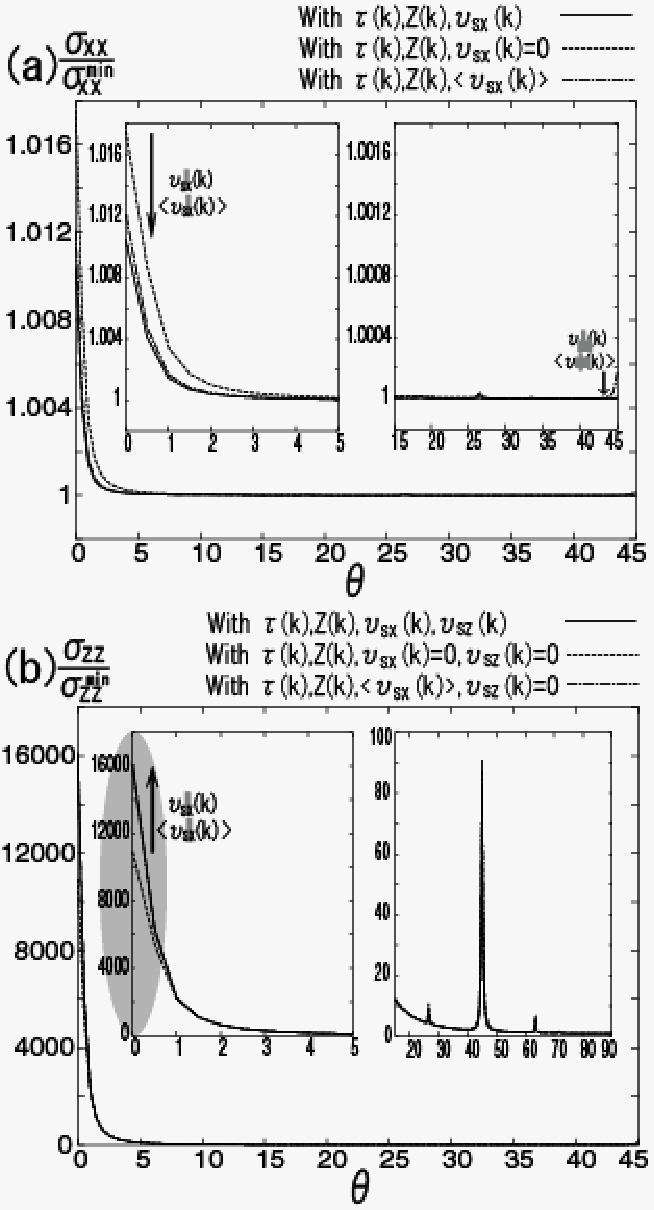}
\end{center}
\caption{(a) Angle dependence of $\sigma_{xx}$. (b) Angle dependence of $\sigma_{zz}$. 
The solid lines mean the magnetoconductivities with using the momentum dependent $\tau({\bf k})$, $Z({\bf k})$ and $\upsilon_{sx}({\bf k})$ (and $\upsilon_{sz}({\bf k})$ in (b)).
The dash lines indicate the magnetoconductivies 
in the case that the momentum dependence of 
$\upsilon_{sx}({\bf k})$ (and $\upsilon_{sz}({\bf k})$ in (b)) is neglected by using $\upsilon_{sx}({\bf k})=0$ (and $\upsilon_{sz}({\bf k})=0$ in (b)). 
The dash line with dot means the magnetoconductivies 
in the case that the constant component of $\upsilon_{sx}({\bf k})$ is treated by using $\langle\upsilon_{sx}({\bf k})\rangle$ (and $\upsilon_{sz}({\bf k})=0$ in (b)).
}
\label{fig3}
\end{figure}
\begin{figure}[!ht]
\begin{center}
\includegraphics[width=0.5\textwidth]{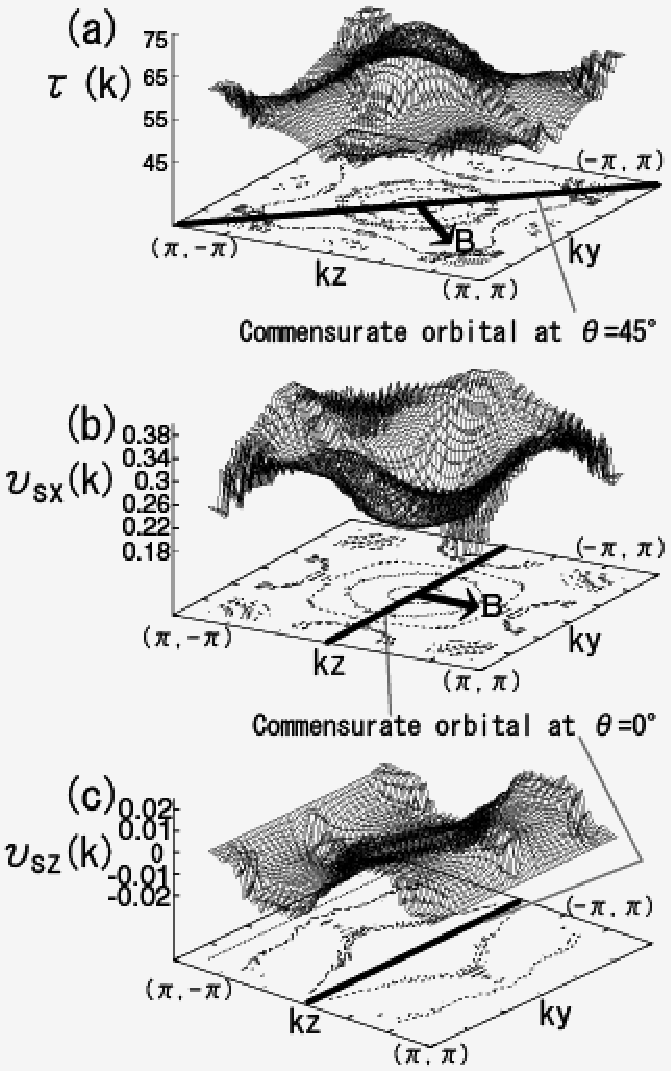}
\end{center}
\caption{(a) $\tau({\bf k})$, (b) $\upsilon_{sx}({\bf k})$ and (c)  $\upsilon_{sz}({\bf k})$ on FS 
and commensurate orbitals at a magic angles $45^\circ$ and $0^\circ$.
}
\label{fig4}
\end{figure}

In our result for the noninteracting system ($U=0$), 
the peaks of $\sigma_{zz}$ appear at magic angles $\theta=0^{\circ}$, $26.6^{\circ}(=\tan^{-1}(\frac{1}{2}))$ and $45^{\circ}(=\tan^{-1}(1))$. 
The angle dependence of $\sigma_{zz}$ in the noninteracting system corresponds to the angle dependence due to the topology of the unlinearized dispersion on FS.\cite{OSADA2} 
At $U=0$, we obtain that the small peaks of $\sigma_{xx}$ also appear at the same magic angles originating from the topology of FS. 
The angle dependences of $\sigma_{xx}$ and $\sigma_{zz}$ are equal to the angle dependence in the non-interacting system 
when we neglect the momentum dependence of $\tau({\bf k})$, $Z({\bf k})$ and $\upsilon_{sx}({\bf k})$ by using $\langle\tau({\bf k})\rangle$, $\langle Z({\bf k})\rangle$, $\upsilon_{sx}({\bf k})=0$ and $\upsilon_{sz}({\bf k})=0$, as shown in Figs.1(a) and 1(b). 

As show in Fig.1(a), 
$\sigma_{xx}$ is strongly suppressed at the magic angle $\theta=45^\circ$ when we calculate it using the momentum dependent $\tau({\bf k})$, $Z({\bf k})$ and  $\upsilon_{sx}({\bf k})$. 
The effect is small at other angles $\theta=0^\circ$ and $26.6^\circ$. 
We calculate $\sigma_{xx}$ by neglecting only the momentum dependence of $\tau({\bf k})$ (using the averaged value  $\langle\tau({\bf k})\rangle$). 
In this case, the peak of $\sigma_{xx}$ at $\theta=45^\circ$ is large as show in Fig. 2(a). 
Therefore, we conclude that the suppression in the peak of $\sigma_{xx}$ at $45^\circ$ due to EC comes through the momentum dependence of $\tau({\bf k})$.
The region of the small value of $\tau({\bf k})$ is attributed to the verge of the SDW. The region of small lifetime originates from the nesting of FS near the SDW in as shown Fig. 4(a).
The effects of the momentum dependence of $\tau({\bf k})$ in $\sigma_{zz}$ are very small as shown in Fig. 2(b). 

Figure 1(b) shows that the peak of $\sigma_{zz}$ is enhanced at $\theta=0^\circ$ when we take the momentum dependences of $\tau({\bf k})$, $Z({\bf k})$, $\upsilon_{sx}({\bf k})$ and $\upsilon_{sz}({\bf k})$ into account. 
The effect of EC in $\sigma_{xx}$ is more sensitive than in $\sigma_{zz}$, as shown in Figs. 1(a) and 1(b).
The effect is small at other angles $\theta=26.4^\circ$ and $45^\circ$. 
In Fig. 3(b), 
the strong enhancement at $\theta=0^\circ$ is obtained in the calculation of $\sigma_{zz}{(\bf k)}$ by neglecting only $\upsilon_{sz}({\bf k})$ and the momentum dependence of $\upsilon_{sx}({\bf k})$ (Using $\upsilon_{sz}({\bf k})=0$ and the averaged value $\langle\upsilon_{sx}({\bf k})\rangle$.), 
while the enhancement is not obtained in the calculation by neglecting only $\upsilon_{sz}({\bf k})$ and $\upsilon_{sx}({\bf k})$ ($\upsilon_{sz}({\bf k})=0$ and $\upsilon_{sx}({\bf k})=0$.).
Therefore, the results indicate that $\sigma_{zz}$ is also enhanced at $\theta=0$ when we calculate it with using only the average $\langle\upsilon_{sx}\rangle$, 
whereas the effect of $\langle\upsilon_{sz}({\bf k})\rangle$ on $\sigma_{zz}$ is very small. 
The enhancement in the peak of $\sigma_{zz}$ at $\theta=0^\circ$ is caused by the increase in velocity due to $\upsilon_{sx}({\bf k})$. 
EC increases the velocity $\upsilon_{x}({\bf k})$. The averaged value $\langle\upsilon_{sx}({\bf k})\rangle$ is 0.3, as shown in Fig. 4(b). 
The effect of $\upsilon_{sz}({\bf k})$ on $\sigma_{zz}$ is very small because $\langle\upsilon_{sz}({\bf k})\rangle$, as shown in Fig. 4(c), equals 0. 

Here, the effects of $\upsilon_{sx}$ suppress the peak at $0^\circ$ in $\sigma_{xx}$ as shown in Fig. 3(a), whereas the momentum dependence of $\tau({\bf k})$ enhances the peak as shown in Fig. 2(a). Therefore, the change of the peak at $0^\circ$ by EC does not appear in the case (Fig.  1(a)) when the both effects exist.  

The reason why $\tau({\bf k})$ suppresses $\sigma_{xx}$ at $\theta=45^\circ$ is explained as follows. 
On the peak of $\theta=45^\circ$ in $\sigma_{xx}$, 
the momentum dependence of $\tau({\bf k})$ over the commensurate orbitals disturbs the momentum dependence of $\upsilon_{x}({\bf k})$ which causes peaks in $\sigma_{xx}$ at magic angles.
The peak at $\theta=45^\circ$ is suppressed by the electron-electron scattering due to the momentum dependence of lifetime, which originates from EC near the SDW, over the commensurate orbital.
 
Next, the enhancement in $\sigma_{zz}$ at $\theta=0^\circ$ by $\langle\upsilon_{sx}({\bf k})\rangle$ is explained as follows. 
The increase of velocity due to $\langle\upsilon_{sx}({\bf k})\rangle$\cite{SHIMAHARA} near the SDW connects to the pseudo gap,  
which indicates the decrease of the density of state ($\frac{1}{\upsilon_{x}({\bf k})}$ in eq. (6)), 
due to spin fluctuation\cite{KINO}. 
The decrease in the density of states suppresses the amplitude of the momentum dependence of $\upsilon_{x}({\bf k})$, 
which changes the overlap of the momentum dependences of $\frac{\upsilon_{z}({\bf k}(0))}{\upsilon_{x}({\bf k}(0))}$ and $\upsilon_{z}({\bf k}(t))$ in eq. (6). 
The decrease in the amplitudes of $\upsilon_{x}({\bf k})$ gives the small effect on the large amplitude of function in the momentum dependence of $\frac{\upsilon_{z}({\bf k}(0))}{\upsilon_{x}({\bf k}(0))}$ corresponding to the large peak at $0^\circ$, 
whereas it gives the large effect on the small amplitude of function corresponding to the small peaks at $26.4^\circ$ and $45^\circ$.  
Therefore, the suppression of the large peak at $0^\circ$ becomes weaker than the suppression of the peaks at the other magic angles of $26.4^\circ$ and $45^\circ$. 

To study the effect of the renormalization factor $Z({\bf k})$,  we calculate $\sigma_{xx}$ and $\sigma_{zz}$ by neglecting only the momentum dependence of $Z({\bf k})$ using $\langle Z({\bf k}) \rangle$. 
We find that the effect of $Z({\bf k})$ is very small in both $\sigma_{xx}$ and $\sigma_{zz}$. The effect of $Z({\bf k})$ is  almost canceled in eqs. (5) and (6).

In comparison with the experimental results, 
we first comment on our results in $\sigma_{zz}$. 
The angular dependence is strong in $\sigma_{zz}$ in the noninteracting system, 
whereas the relative height of peaks between magic angles does not agree with the experimental results.
The effect of EC changes the relative height of the peaks in $\sigma_{zz}$, 
however the change is not sufficient to explain the experimental results. 
Next, we comment on $\sigma_{xx}$. 
In the Chaikin's simple model\cite{CHAIKIN} of "hot spot",  
the strong angular dependence is obtained in 1D axis magnetoresistance, 
as observed experimentally. 
Zhelenyak and Yakovenko\cite{YAKOVENKO} have found that the variation of the lifetime on the Q1D FS calculated on the basis of Umklapp  scattering is not strong enough to give the peaks in the 1D axis magnetoresistance 
for the model with linearized dispersion in the $k_x$-direction. 
In the present study, 
we did not linearize the dispersion 
and we found that EC changes the relative height of the peaks in the angular dependence of $\sigma_{xx}$. 
The heights of peaks, however, are not as high as those observed experimentally.
One of the possible origins of the strong angle dependence may be the vertex corrections. 
Kontani\cite{KONTANI} has shown that vertex correction plays an important role in the magnetoresistance of two-dimensional systems in the weak magnetic field. 
By using the quantum Kubo formula, the vertex correction should be taken into account in general, however, no formulation in the strong magnetic field has been obtained, as far as the authors know. 

The Zeeman effect is not considered in this study. 
It will also affect the result, 
because $\chi_{\uparrow \uparrow}$, $\chi_{\downarrow \downarrow}$ and $\chi_{\uparrow \downarrow}$ are not the same in the presence of the Zeeman effect.

In summary, the momentum dependence of the quasi-particle's lifetime originating from EC near the SDW, which causes electron-electron scattering, suppresses $\sigma_{xx}$ at the magic angle of $45^\circ$.
The change of the quasi-particle's velocity due to EC, which connects to the pseudo gap due to spin fluctuation near the SDW, gives the enhancement in $\sigma_{zz}$ at the magic angle $0^\circ$. 
The different effects of EC between the magic angles originate from the connection of the momentum dependences of lifetime and velocity with the momentum dependence of commensurate orbitals on a Fermi surface in magnetic fields at magic angles. The effect of EC is sensitive to $\sigma_{xx}$. The effect of EC on AMRO may be considered as the origin of the deviation of the peak heights obtained in the experiments from that calculated in the noninteracting system. 

One of the authors (H. K.) thank Professor K. Makosi 
for his support and encouragement. 
H. K would like to thank Professor H. Kontani 
for his valuable discussions. 
H. K. thanks Dr. T. Jujo and Dr. Y. Yanase 
for their illuminate comments. 
This work was supported by Scientific
Research on Priority Area ''Novel Functions of Molecular Conductors Under Extreme Conditions'' under the Grants-in-Aid for Scientific Research of the Japan Society for the Promotion of Science.


\begin{thebibliography}{99}
\bibitem{FISDW1}
D. Jerome and H. J. Schulz: Adv. Phys. \textbf{31} (1982)  299.
\bibitem{FISDW2}
T. Ishiguro, K. Yamaji and G. Sato: Organic Superconductors, Springer-Verlag, Berlin (1998).
\bibitem{AMRO1}
M. J. Naughton, O. H. Chung and M. Chaparala: Phys. Rev. Lett. \textbf {67} (1991) 3712.
\bibitem{AMRO2}
T. Osada, A. Kawasumi, S. Kagoshima, N. Miura, and G. Saito: Phys. Rev. Lett. \textbf{66} (1991) 1525.
\bibitem{AMRO3}
W. Kang, S. T. Hannas and P. M. Chaikin: Phys. Rev. Lett. \textbf{69} (1992) 2827.
\bibitem{OSADA1} 
T. Osada, S. Kagoshima and N. Miura: Phys. Rev. B. 
\textbf{46} (1992) 1812.
\bibitem{OSADA2} 
T. Osada, S. Kagoshima and N. Miura: Phys. Rev. Lett. 
\textbf{77} (1996) 5261.
\bibitem{MAKI}
K. Maki: Phys. Rev. B \textbf{45} (1992) 5111.
\bibitem{HASEGAWA}
Y. Hasegawa, H. Kaneyasu and K. Kishigi:J. Phys. Soc. Jpn. \text{75} (2006) No.4, In press.
\bibitem{UMKLAPP}
L. B. Ioffe and A. J. Millis: Phys Rev. B \textbf{58} (1998) 11631.
\bibitem{LEBED}
A. G. Lebed and P. Bak: Phys. Rev. Lett. \textbf{63} (1989) 1315.
\bibitem{CHAIKIN}
P. M. Chaikin: Phys. Rev. Lett. \textbf{69} (1992) 2831.
\bibitem{YAKOVENKO}
A. T. Zheleznyak and V. M. Yakovenko: Synthetic Metals \textbf{70} (1995) 1005.
\bibitem{MCKENZIE}
P. Moses and R. H. McKenzie: Phys. Rev. B \textbf{63} (2000) 024414.
\bibitem{RPA1}
S. Takada: J. Phys. Soc. Jpn. \textbf{53} (1984) 2193.
\bibitem{RPA2}
K. Maki and A. Virosztek: Phys. Rev. B \textbf{36} (1987) 511.
\bibitem{YANASE}
Y.Yanase and K.Yamada: J. Phys. Soc. Jpn. \textbf{68} (1999) 548.
\bibitem{ASCROFT} 
N. W. Ashcroft and N. D. Mermin: Solid State Physics (Saunders, Philadelphia) (1975).
\bibitem{SHIMAHARA}
H.Shimahara: J. Phys. Soc. Jpn. \textbf{58} (1989) 1735.
\bibitem{KINO}
H. Kino and H.Kontani: J. Phys. Soc. Jpn. \textbf{68} (1999) 1481.
\bibitem{KONTANI}
H. Kontani: J. Phys. Soc. Jpn. \textbf{68} (1999j614.
\end{thebibliography}
\end{document}